\documentclass[amssymb,amsmath,aps,groupedaddress,reprint,superscriptaddress,showpacs]{revtex4}

\usepackage{graphicx}
\usepackage{dcolumn}
\usepackage{bm}
\usepackage{amsmath}

\begin{document}

\title{Valley entanglement of excitons in monolayers of transition-metal dichalcogenides}

\author{Mikhail Tokman}
\affiliation{Institute of Applied Physics, Russian Academy of Sciences, 46 Ulyanov Street , 603950 Nizhny Novgorod , Russia }
\author{Yongrui Wang}
\affiliation{Department of Physics and Astronomy, Texas A\&M
University, College Station, TX, 77843 USA}
\author{Alexey Belyanin}
\email{belyanin@tamu.edu}
\affiliation{Department of Physics and Astronomy, Texas A\&M
University, College Station, TX, 77843 USA}

\begin{abstract}
We show that excitons and free carriers in K and K' valleys of transition metal dichalcogenide monolayers can be entangled with respect to their valley degree of freedom by absorbing linearly polarized single photons. This effect does not require any interaction between K and K' excitons in contrast to conventional mechanisms of entanglement that are mediated by coupling between quantum systems (e.g.~entanglement of photons in nonlinear optical interactions). The valley entanglement of excitons and free carriers can be verified by measuring the polarization of their photoluminescence or  fluctuations of the photocurrent under an applied in-plane DC bias.

\end{abstract}

\date{\today}


\maketitle

\section{Introduction}

Transition metal dichalcogenides (TMDCs) of the composition MX$_2$, where M = Mo or W and X = S, Se, or Te, are well known  materials that have recently seen the resurgence of interest. This interest was largely driven by two discoveries. First, although bulk MoS$_2$ is an indirect-gap material, the monolayer MoS$_2$ turns out to have the direct band gap located in energy-degenerate $K$ and $K'$ points at the corners of the hexagonal Brillouin zone. This results in a dramatic increase of the band-edge photoluminescence (PL) yield by more that a factor of $10^4$ \cite{mak2010}. Second, the combination of spatial inversion symmetry breaking with strong spin-orbit coupling leads to a valley-contrasting spin splitting of the valence-band edge, which gives rise to valley-dependent optical selection rules for interband transitions \cite{xiao2012}.  Namely, free carriers and excitons in the $K$ and $K'$ valleys are coupled to photons of the same energy but opposite helicity: left-hand circular (L) polarization for the $K$ point and right-hand circular (R) polarization for the $K'$ point. When excited by a circularly polarized light,  MoS$_2$ monolayer shows PL with the same polarization as the excitation light, indicating that the valley polarization of excitons is preserved longer than the recombination time, at least at low temperatures \cite{zeng2012, mak2012,cao2012}. When excited by a linearly polarized light, WSe$_2$ shows a high degree of linear polarization in the PL of neutral excitons, indicating that the inter-valley phase coherence survives in the processes of exciton formation and recombination \cite{jones2013}. Under an applied bias, valley-polarized TMDC monolayers exhibit valley and spin Hall effects \cite{xiao2012} which enables optoelectronic devices based on the valley degree of freedom, e.g. valley Hall effect transistors  \cite{mak2014}.  

These recent results are very exciting as they suggest that the valley index degree of freedom in TMDC monolayers  can serve as a robust information carrier. Excitons or free carriers in different valleys can be selectively manipulated by radiation in the convenient visible frequency range around 2 eV and by the in-plane DC electric field. A tantalizing question with implications for quantum information is whether and how one can achieve quantum mechanical entanglement of free carriers or excitons with respect to the valley index.

The entanglement of two quantum systems or ensembles is usually generated as a result of coupling between them. This coupling can be mediated by classical electromagnetic fields; see e.g. \cite{scully,mikhail,vdovin}. However,  a classical field cannot entangle {\it non-interacting} systems unless it directly couples them through a nonlinear optical process such as parametric frequency conversion or four-wave mixing. For example, the experiment in \cite{jones2013} which employed a classical field cannot lead to entanglement of excitons (see Appendix A).

At the same time, one can entangle non-interacting quantum systems by coupling them to a {\it quantum} field. 
Here we consider the optical excitation of electron-hole or neutral exciton states near the band gap of a MX$_2$ monolayer in two valleys K and K' with opposite valley indices. We will consider entanglement of excitons for definiteness, although our analysis below works  equally well for free electron-hole pairs. The binding energy of excitons in MX$_2$ monolayers is as high as several hundred meV, and the PL signal is dominated by excitons even for above band gap excitation. We will label two different valley states by an up or down pseudospin direction, assuming that ($\uparrow$) and ($\downarrow$) excitons can be excited only by an R- or L-polarized field, respectively:
$$
\textbf{E}_{R,L} \propto \textbf{e}_{\pm} = \frac{\textbf{x} \pm i\textbf{y}}{\sqrt{2}}
$$

\begin{figure}[htbp]
\includegraphics[width=0.5\textwidth]{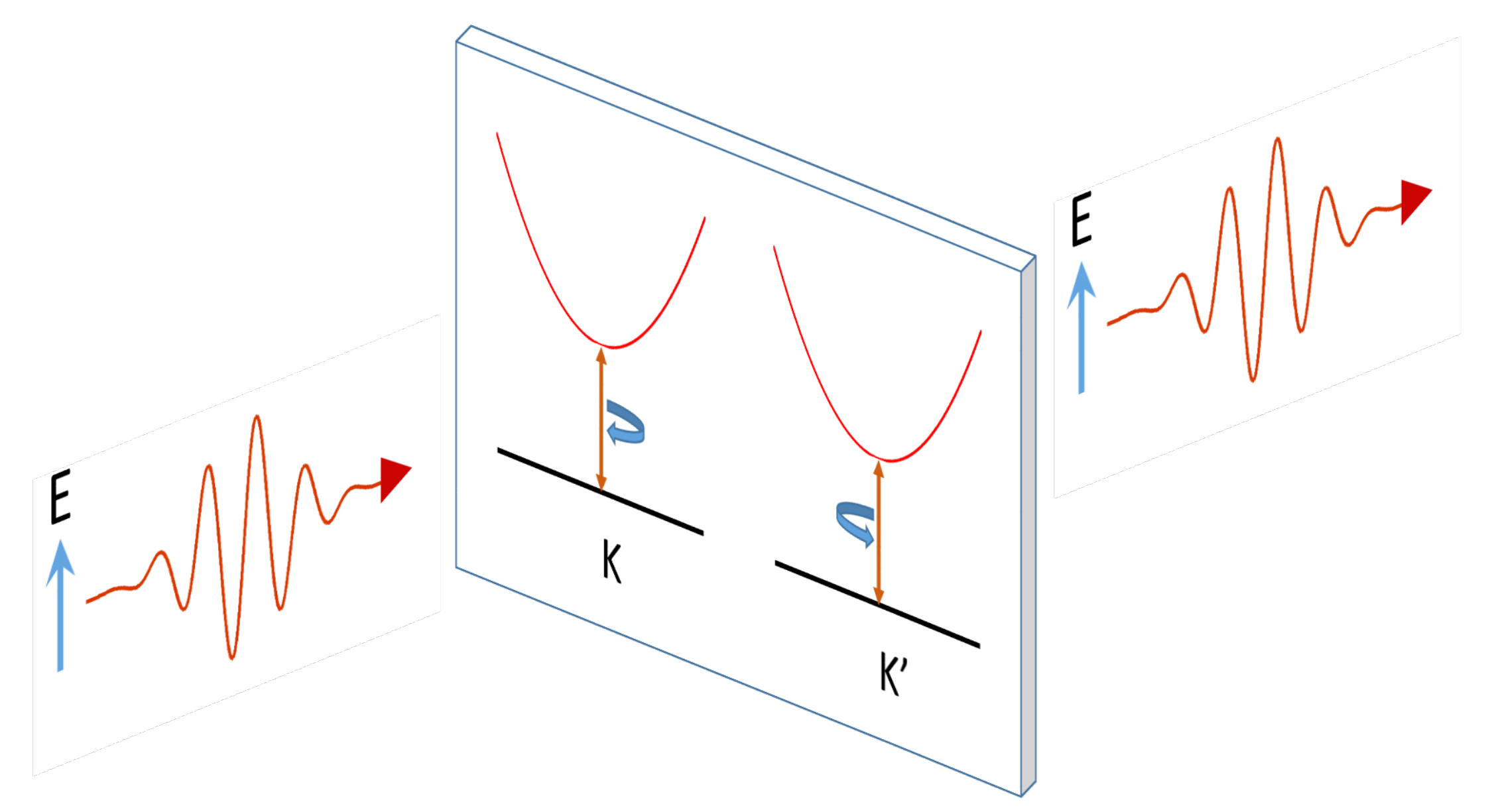}
\caption{(Color online) A sketch of excitons in K and K' valleys of MX$_2$ monolayers interacting with linearly polarized single photons. To ensure the interaction with one photon at a time, the power of incident radiation should not exceed $\hbar \omega \Delta \omega$, where  $\Delta \omega$ is the radiation bandwidth. }
\end{figure}


The main result of the paper is that absorption of linearly polarized single photons by a MX$_2$ monolayer in a cavity (Sec.~II)  or from a flux of incident photons (Sec.~III)  gives rise to an efficient entanglement of up and down excitons, i.e. of the valley degree of freedom; see Fig.~1. The intuitive physical explanation of this effect is that excitons in the K and K' valleys of a MX$_2$ monolayer interact with a linearly polarized single photon as if it were an entangled R/L photon state. In fact, one can rigorously prove (see Appendix B) that a single photon state which is a factorized product state in a linearly polarized basis is equivalent to an entangled state in the basis of R and L photon modes. As a result, absorption of such a photon leads to valley entanglement of excitons or free carriers. 

Valley entanglement of photoexcited excitons can be verified by measuring the polarization of their PL (Sec.~III) or  fluctuations of the photocurrent under an applied in-plane DC bias (Sec.~IV).

\section{Entanglement of the excitonic ensemble in a cavity}

To visualize the effect of entanglement, here we consider a toy problem keeping only the light-matter interaction and neglecting all processes leading to decoherence and losses. In the next section we will include incoherent processes, finite bandwidth, fluctuations, and losses. 

\begin{figure}[htbp]
\includegraphics[width=0.5\textwidth]{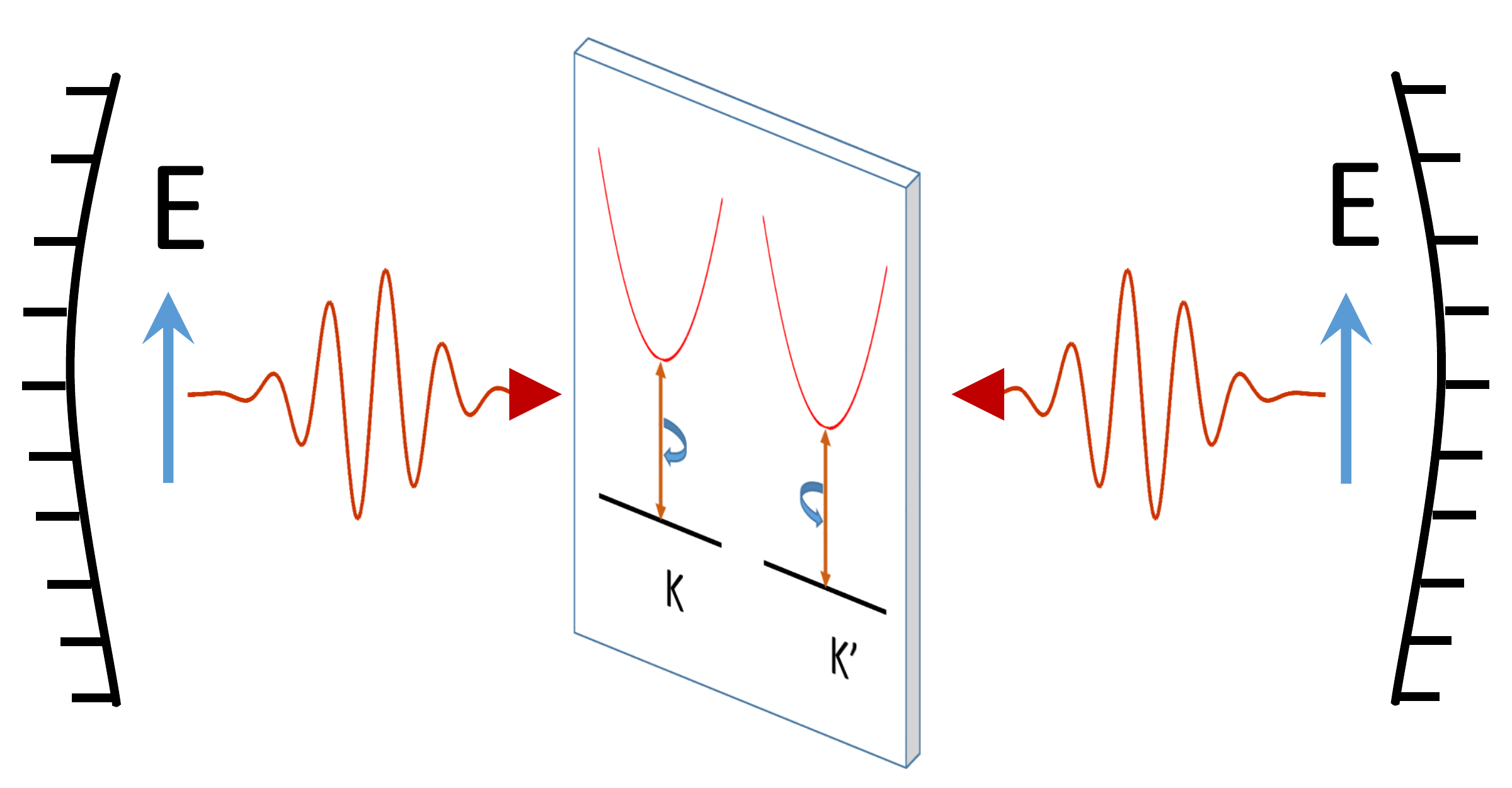}
\caption{(Color online) Valley entanglement of excitons in K and K' valleys of a MX$_2$ monolayer in a cavity geometry. }
\end{figure}


Consider a MX$_2$ monolayer placed into a cavity which supports identical R and L modes of the same photon energy $\hbar\omega$ equal to the exciton energy, see Fig.~2. The quantized cavity field is given by
\begin{eqnarray}
\label{cavity} 
&&\hat{\textbf{E}} = \textbf{E}_R(\textbf{r})\hat{c}_R + \textbf{E}^*_R(\textbf{r})\hat{c}^{\dag}_R + \textbf{E}_L(\textbf{r})\hat{c}_L + \textbf{E}^*_L(\textbf{r})\hat{c}^{\dag}_L,\nonumber\\
&&\textbf{E}_{R,L}(\textbf{r}) = \textbf{e}_{\pm}E_{R,L}(\textbf{r}), |E_R(\textbf{r})|=|E_L(\textbf{r})|= E(\textbf{r}),\nonumber\\
\end{eqnarray}
where the spatial mode distributions are normalized according to
\begin{equation}
\label{norm}
\int_V\varepsilon(\textbf{r})E^2(\textbf{r})d^3r = 4\pi\hbar\omega,
\end{equation}
where $\varepsilon(\textbf{r})$ is the intracavity distribution of the dielectric constant, which may include also the monolayer substrate and is assumed to be frequency-independent for simplicity. It can be easily generalized to dispersive media, see e.g. \cite{mikhail,vdovin}. Note that the normalization (\ref{norm}) takes care of both the electric and magnetic field energy, because for stationary fields in a cavity or for periodic boundary conditions one can prove
$
\int_V B^2(\textbf{r})d^3r = \int_V\varepsilon(\textbf{r})E^2(\textbf{r})d^3r.
$
The field satisfying Eqs.~(\ref{cavity}) and (\ref{norm}) corresponds to the Hamiltonian
\begin{equation}
\hat{H}_f = \hat{H}_R + \hat{H}_L = \hbar\omega \hat{c}^{\dag}_R\hat{c}_R 
+ \hbar\omega \hat{c}^{\dag}_L\hat{c}_L ,
\end{equation}
where photon creation and annihilation operators $\hat{c}^{\dag}_R, \hat{c}^{\dag}_L$ and  $\hat{c}_R, \hat{c}_L$  satisfy standard commutation relations \cite{scully}. We will assume the fields $\textbf{E}_{R,L}(\textbf{r})$ and $\textbf{E}_{R,L}(\textbf{r})$ to be uniform on the monolayer and omit the position argument.

We will treat the radiation as quasi-monochromatic and assume that there are a total of $N$ pairs of up and down exciton states in a monolayer of area S that can be excited by photons of a given energy. More precisely, $N$ is a number of electron-hole or exciton states within the interband transition linewidth. For a broadband radiation $N$ would be a number of states within the radiation bandwidth.

To simplify the notations, we will denote ($\uparrow$) excitons with latin indices and ($\downarrow$) excitons with greek indices. The Hamiltonian of an ensemble of excitons is given by
\begin{equation}
\hat{H}_e = \hat{H}_{\uparrow} + \hat{H}_{\downarrow} = \hbar\omega\sum_{j=1}^N|1_j\rangle\langle1_j|
+ \hbar\omega\sum_{\xi=1}^N|1_{\xi}\rangle\langle1_{\xi}|,
\end{equation}
where the ground state $|0_{j,\xi}\rangle$(no exciton) corresponds to zero energy, whereas the state $1_{j,\xi}$ describes an excited exciton of energy $\hbar\omega$. The polarization operator of the system can be written as
\begin{eqnarray}
\hat{\textbf{P}}_e &=& \sum^N_{j=1}\left( \textbf{d}_{\uparrow}|1_j\rangle\langle0_j| + \textbf{d}^*_{\uparrow}|0_j\rangle\langle1_j| \right)\nonumber\\
 &+&  \sum^N_{\xi=1}\left( \textbf{d}_{\downarrow}|1_{\xi}\rangle\langle0_{\xi}| + \textbf{d}^*_{\downarrow}|0_{\xi}\rangle\langle1_{\xi}| \right),
\end{eqnarray}
where the dipole moments
$$
\textbf{d}_{\uparrow} \equiv \textbf{d}_{(10)\uparrow} = {\bf e_-}d, \textbf{d}_{\downarrow} \equiv \textbf{d}_{(10)\downarrow}=\textbf{d}_{(01)\uparrow}\equiv \textbf{d}^*_{\uparrow}= {\bf e_+}d^*.
$$

The Hamiltonian describing interaction between the field and the particles in the electric dipole approximation is
\begin{equation}
\hat{V} = -\hat{\textbf{P}}_e(\textbf{E}_R\hat{c}_R + \textbf{E}^*_R\hat{c}^{\dag}_R +\textbf{E}_L\hat{c}_L + \textbf{E}^*_L\hat{c}^{\dag}_L ).
\end{equation}
Consider Schroedinger's equation
\begin{equation}
\label{schrod}
i\hbar\dot{\Psi} = (\hat{H}_R + \hat{H}_L + \hat{H}_{\uparrow} + \hat{H}_{\downarrow} +\hat{V})\Psi.
\end{equation}
We choose the following initial state as a product of the excitonic, $\Psi_e(0)$, and field, $\Psi_f(0)$ wave functions:
\begin{eqnarray}
&&\Psi(0) = \prod^N_{j=1}|0_j\rangle\prod^N_{\xi=1}|0_{\xi}\rangle(C_R(0)|1_R\rangle|0_L\rangle \nonumber\\
 &&+ C_L(0)|1_L\rangle|0_R\rangle),\nonumber\\
&&|C_R(0)|^2 + |C_L(0)|^2=1
\end{eqnarray}
in which excitons are not excited and the single-photon field has an arbitrary elliptical polarization. The particular case of a linear polarization corresponds to $|C_R|^2=|C_L|^2$(see Appendix B). 

The state at an arbitrary moment of time at the same energy $\hbar\omega$ which is conserved as a result of field-matter interaction is
\begin{eqnarray}
\label{initial}
\Psi &=& \sum_{j = 1}^N C_j(t) |1_j\rangle\prod^N_{i\ne j}|0_{i}\rangle\prod^N_{\xi=1}|0_{\xi}\rangle|0_R\rangle|0_L\rangle e^{-i\omega t}\nonumber\\
&+&  \sum^N_{\xi=1} C_{\xi}(t) |1_{\xi}\rangle\prod^N_{\eta\ne\xi}|0_{\eta}\rangle\prod^N_{j=1}|0_{j}\rangle|0_R\rangle|0_L\rangle e^{-i\omega t}\nonumber\\
&+&  C_R(t)|1_R\rangle|0_L\rangle\prod^N_{j=1}|0_j\rangle\prod^N_{\xi=1}|0_{\xi}\rangle e^{-i\omega t}\nonumber\\
&+& C_L(t)|1_L\rangle|0_R\rangle\prod^N_{j=1}|0_j\rangle\prod^N_{\xi=1}|0_{\xi}\rangle e^{-i\omega t}.
\end{eqnarray}
Substituting (\ref{initial}) into (\ref{schrod}), we obtain the equations for coefficients
\begin{equation}
\label{amp}
\left\{\begin{array}{cc}
i\hbar\dot{C}_R = -\left(\textbf{d}_{\uparrow}\textbf{E}^*_R\right)\sum^{N}_{j=1}C_j,&
i\hbar\dot{C}_j = -\left(\textbf{d}_{\uparrow}\textbf{E}_R\right)C_R \\ \\
i\hbar\dot{C}_L = -\left(\textbf{d}_{\downarrow}\textbf{E}^*_L\right)\sum^{N}_{\xi=1}C_{\xi},&
i\hbar\dot{C}_{\xi} = -\left(\textbf{d}_{\downarrow}\textbf{E}_L\right)C_L
\end{array}\right..
\end{equation}
After introducing the complex Rabi frequency $\displaystyle \frac{\textbf{d}_{\uparrow}\textbf{E}^*_R}{\hbar} = \Omega = |\Omega|e^{i\Theta}$, Eqs.~(\ref{amp}) can be written as
\begin{equation}
\left\{\begin{array}{cc}
i\dot{C}_R = -\Omega\sum^{N}_{j=1}C_j,&
i\dot{C}_j = -\Omega^*C_R \\ \\
i\dot{C}_L = -\Omega^*\sum^{N}_{\xi=1}C_{\xi},&
i\dot{C}_{\xi} = -\Omega C_L
\end{array}\right..
\end{equation}
Taking the second derivative yields
\begin{equation}
\ddot{C}_{R,L} + N|\Omega|^2C_{R,L} = 0.
\end{equation}
For the initial field in the linearly polarized product state, which is the Bell state (\ref{bell}) in the circularly polarized basis (see Appendix B), i.e. when
\begin{equation}
\label{bell}
\begin{array}{c}
C_R(0) = \displaystyle \frac{1}{\sqrt{2}}, C_L(0) = \pm\frac{1}{\sqrt{2}}\\
\Psi_f(0) = \Psi_{\pm} = \displaystyle  \frac{|1_R\rangle|0_L\rangle\pm|1_L\rangle|0_R\rangle}{\sqrt{2}}
\end{array}
\end{equation}
we obtain the solution
\begin{equation}
\label{sol-bell}
C_R = \frac{\cos{(\sqrt{N}|\Omega|t)}}{\sqrt{2}}, C_L = \pm\frac{\cos{(\sqrt{N}|\Omega|t)}}{\sqrt{2}}.
\end{equation}
Substituting (\ref{sol-bell}) into (\ref{amp}), we can obtain for $C_{j,\xi}$:
\begin{equation}
\label{sol-exc}
C_j = -i\frac{e^{-i\Theta}}{\sqrt{2N}}\sin{(\sqrt{N}|\Omega|t)},
C_{\xi} = \mp i\frac{e^{i\Theta}}{\sqrt{2N}}\sin{(\sqrt{N}|\Omega|t)}.
\end{equation}

Therefore, the solution of Eq. (\ref{schrod}) when the field is initially in the Bell state (\ref{bell}) and excitons are in the ground state has the form
\begin{eqnarray}
\label{sol-sch}
&&\Psi = e^{-i\omega t}\prod^N_{j=1}|0_j\rangle\prod^N_{\xi=1}|0_{\xi}\rangle\frac{\cos{(\sqrt{N}|\Omega|t)}}{\sqrt{2}}
\left( |1_R\rangle|0_L\rangle\right. \nonumber\\
&&\pm \left.|0_R\rangle|1_L\rangle \right) \pm ie^{-i\omega t} |0_R\rangle|0_L\rangle  \frac{\sin{(\sqrt{N}|\Omega|t)}}{\sqrt{2N}}\left(e^{-i\Theta}\sum^N_{j=1}|1_j\rangle\right.\nonumber\\
&&\left.\prod^N_{i\ne j}|0_{i}\rangle\prod^N_{\xi=1}|0_{\xi}\rangle\pm  e^{i\Theta}\sum^N_{\xi=1}|1_{\xi}\rangle\prod^N_{\eta\ne\xi}|0_{\eta}\rangle\prod^N_{j=1}|0_j\rangle\right).
\end{eqnarray}
For $\sqrt{N}|\Omega|t=\frac{\pi}{2}$  the state (\ref{sol-sch}) corresponds to the Bell-type entangled state of excitons:
\begin{eqnarray}
\label{sol-sch2}
\Psi &=& \frac{|0_R\rangle|0_L\rangle}{\sqrt{2N}}
\left(e^{-i\Theta}\sum^N_{j=1}|1_j\rangle
\prod^N_{i\ne j}|0_{i}\rangle\prod^N_{\xi=1}|0_{\xi}\rangle \right.\nonumber\\
&\pm&  \left. e^{i\Theta}\sum^N_{\xi=1}|1_{\xi}\rangle\prod^N_{\eta\ne\xi}|0_{\eta}\rangle\prod^N_{j=1}|0_j\rangle\right)
\end{eqnarray}

This result is intuitively expected: the excitons in each valley couple directly to only R- or L-component of the linearly polarized photon state, but these components  were entangled with each other, so the entanglement passes on to the excitonic system. Clearly this mechanism of entanglement exists only for a quantum incident field. Since there is no decoherence or loss, the energy and entanglement oscillate back and forth between the excitons and the photon field. If at time t = 0 the cavity contained exactly one photon of a linearly polarized field, then after the time $t=\frac{\pi}{2\sqrt{N}|\Omega|}$ it will be absorbed by an ensemble of excitons. Since the photon energy is equal to the energy of one exciton, one could (wrongly) assume that this photon creates one (up or down) exciton which will eventually recombine, so the reemitted photon will be in either R or L state with equal probability. However, from the exact solution (\ref{sol-sch}) we see that the reemitted photon will be linearly polarized. Interestingly, it was predicted in \cite{scully2006} that the spatial structure of a single-photon field should be preserved after its absorption and reemission by an ensemble of atoms due to the entanglement of their states. Here we obtain a conceptually similar result for the polarization of the field.

We also note for the subsequent discussion that the state (\ref{sol-sch2}) satisfies the following condition for any pair of ($\uparrow$) and ($\downarrow$) excitons:
\begin{equation}
\label{pair}
\langle \hat{\rho}_{(11)j}\hat{\rho}_{(11)\xi}\rangle = 0,
\end{equation}
where $\hat{\rho}_{(11),j}, \hat{\rho}_{(11),\xi}$ are operators of the upper state population: $\hat{\rho}_{(11)j} = |1_j\rangle\langle1_j|$, $\hat{\rho}_{(11)\xi} = |1_{\xi}\rangle\langle1_{\xi}|$.


\section{Entanglement of the excitonic ensemble in a transmission geometry}


\subsection{The Heisenberg-Langevin formalism}

In this section we consider the response of excitons in a MX$_2$ monolayer to an illumination by a stationary flux of photons, including the effects of relaxation, losses, finite spectral bandwidth, and fluctuations, having in mind a generic geometry of Fig.~1. We will solve the Heisenberg-Langevin equation for the density operator of an ensemble of ($\uparrow$) or ($\downarrow$) excitons (see e.g. \cite{mikhail,JL}):
\begin{equation}
\label{master}
\frac{\partial \hat{\rho}_{mn}}{\partial t} = \frac{-i}{\hbar}
\left(\hat{h}_{mp}\hat{\rho}_{pn} - \hat{\rho}_{mp}\hat{h}_{pn}
\right) + \hat{R}_{mn} + \hat{F}_{mn}.
\end{equation}
Here $\hat{\rho}_{mn}(\textbf{r},t)$ is the Heisenberg density operator, $\hat{R}_{mn}$ the relaxation operator, $\hat{F}_{mn}$ is the Langevin noise operator describing fluctuations in an excitonic system. The Heisenberg density matrix operator can be determined in a number of ways. One can use the projection operator $\hat{\rho}_{mn} = | n\rangle \langle m |$ \cite{scully,loudon}. Another way is to use operators of annihilation and creation of quantum states,  $\hat{\rho}_{mn} = \hat{a}_n^{\dagger} \hat{a}_m$ \cite{scully,haken}. Both approaches lead to quite similar results. We will call the operator matrix $\hat{\rho}_{mn} $  a density operator regardless of its representation. The Hamiltonian is
\begin{equation}
\label{ham}
\hat{h}_{mn} = W_n\delta_{nm} - \textbf{d}_{nm}\hat{\textbf{E}}(z=0),
\end{equation}
where the monolayer is located in the z=0 plane, $W_n$ are eigen values of the unperturbed Hamiltonian of excitons. The density operator determined from (19) allows one to calculate the spatial density of any physical quantity $A(\textbf{r},t)$:
$$
A(\textbf{r},t) = \langle\Psi_0 |A_{nm}\hat{\rho}_{mn}(\textbf{r},t) |\Psi_0 \rangle,
$$
where $A_{mn}$ is the matrix element of the operator $\hat{A}$ corresponding to quantity A and $|\Psi_0\rangle$ is the initial state function in the Heisenberg picture. If the initial wave function is normalized to unity, $\langle\Psi_0|\Psi_0\rangle = 1$, the normalization of the density operator of an extended system is $\sum_n\hat{\rho}_{nn}(\textbf{r},t) = N(\textbf{r})$, where $N(r)$ is the density of particles which is assumed to be constant in this case. For a monolayer the quantity $N(\textbf{r}) = N(\textbf{r}_{\perp})$ is a surface density, where $\textbf{r}_{\perp}$ is the radius-vector in the monolayer plane.

The Langevin noise operator satisfies the general conditions $\hat{F}_{nm} = \hat{F}^{\dag}_{mn}$ , $\langle\hat{F}_{mn}\rangle = 0$ , where $\langle...\rangle$ means averaging over the statistics of a noise reservoir. Other properties of the noise operator are determined by the properties of the relaxation operator. We will define the latter in the simplest form through a constant relaxation rate:
\begin{equation}
\label{relax}
\left\{\begin{array}{c}
\hat{R}_{mn} = -\gamma_{mn}\hat{\rho}_{mn}, \quad m\ne n \\
\hat{R}_{mm} = \sum_n w_{mn}\hat{\rho}_{nn}\end{array}
\right.
\end{equation}
The definition (\ref{relax}) implies that the relaxation operator is delta-correlated in time. Using generalized Einstein relations \cite{cohen,scully,sargent} and assuming also delta-correlation in space, one can derive the following relationship for the noise correlators for a two-level system \cite{tokman,david}:
\begin{eqnarray}
\label{corr}
&&\langle \hat{F}_{mn}^{\dag}(t,\textbf{r}_{\perp}) \hat{F}_{mn}(t',\textbf{r}'_{\perp})\rangle =
\langle 2\gamma_{mn}\hat{\rho}_{mm} + \hat{R}_{mm}\rangle \delta(t-t')\delta(\textbf{r}_{\perp} - \textbf{r}'_{\perp})\nonumber\\
&&\langle \hat{F}_{mn}(t,\textbf{r}_{\perp}) \hat{F}_{mn}^{\dag}(t',\textbf{r}'_{\perp})\rangle =
\langle 2\gamma_{mn}\hat{\rho}_{nn} + \hat{R}_{nn}\rangle \delta(t-t')\delta(\textbf{r}_{\perp} - \textbf{r}'_{\perp}).
\end{eqnarray}

In a two-level system, we denote the transverse relaxation rate as $\gamma_{mn}\equiv\gamma_{10}=\gamma$. Assuming that the exciton system is not too far from the equilibrium and the Fermi level is far below the exciton band edge (so that the number of excitons is always small), we can also obtain
\begin{equation}
\hat{R}_{11} = -\hat{R}_{00} \approx -\Gamma\hat{\rho}_{11}.
\end{equation}
Under these conditions and for $\gamma \gg \Gamma$ we can also take $\hat{R}_{mm}\approx 0$ and $\hat{R}_{nn}\approx 0$ in Eqs.~(\ref{corr}) and (\ref{spectrum}) as in \cite{mikhail}.


\subsection{Response of the MX$_2$ monolayer to an incident field}

Now we derive the optical-frequency current induced in the valleys and the properties of the reemitted field. We start from the following ansatz for the field operator in the monolayer $\hat{\bf E}(z = 0) \equiv \hat{\textbf{E}}(0)$:
\begin{equation}
\label{field}
\hat{\textbf{E}}(0) = \hat{\textbf{E}}^{(+)}e^{-i\omega t} + \hat{\textbf{E}}^{(-)}e^{i\omega t}; \; 
\hat{\textbf{E}}^{(+)} = \textbf{e}_+\hat{E}_R + \textbf{e}_-\hat{E}_L, \; \hat{\textbf{E}}^{(-)} = \left(\hat{\textbf{E}}^{(+)} \right)^{\dag}. 
\end{equation}

Next we will use the equations for the density operator (\ref{master}) and (\ref{ham}). They can be split into two pairs of equations: for "$\uparrow + R$" and "$\downarrow + L$" states. These pairs can be coupled via an entangled initial state as we showed in the previous section; however, this does not affect the form of the equations.

Introducing slowly varying amplitudes of the off-diagonal density operators (''coherences''),
\begin{equation}
\label{slow} 
\hat{\rho}_{(10)\uparrow,\downarrow} = \hat{\sigma}_{(10)\uparrow,\downarrow}e^{-i\omega t},
\hat{\rho}_{(01)\uparrow,\downarrow} = \hat{\sigma}_{(01)\uparrow,\downarrow}e^{i\omega t},
\end{equation}
we obtain the density operator for an R-and L-polarized surface current:
\begin{eqnarray}
\label{current2}
\hat{\textbf{j}}_{\uparrow,\downarrow} &=& (\textbf{e}_{\pm}\hat{j}_{\uparrow,\downarrow}e^{-i\omega t} + \textbf{e}_{\mp}\hat{j}^{\dag}_{\uparrow,\downarrow}e^{i\omega t})\nonumber\\
&=& (\textbf{e}_{\pm}j^*_{\uparrow,\downarrow}\hat{\sigma}_{(10)\uparrow,\downarrow}e^{-i\omega t} + \textbf{e}_{\mp}j_{\uparrow,\downarrow}\hat{\sigma}_{(10)\uparrow,\downarrow}e^{i\omega t}),
\end{eqnarray}
where we introduced the notation $j_{(10)\uparrow}\equiv j_{\uparrow} = i\omega d$, $j_{(10)\downarrow}\equiv j_{\downarrow} = i\omega d^*$.

Assuming that the spectral bandwidth of the radiation does not exceed the relaxation rate $\gamma$  we obtain the stationary solution of Eqs.~(\ref{master}), (\ref{slow}), (\ref{current2}):
\begin{eqnarray}
\label{stat}
&&\hat{j}_{\uparrow} = \frac{\omega|d|^2}{\hbar\gamma}(\hat{\rho}_{(00)\uparrow} - \hat{\rho}_{(11)\uparrow})
\hat{E}_R  - \frac{i\omega d^*}{\gamma}\hat{\tilde{F}}_{\uparrow},\nonumber\\
&&\hat{j}_{\downarrow} = \frac{\omega|d|^2}{\hbar\gamma}(\hat{\rho}_{(00)\downarrow} - \hat{\rho}_{(11)\downarrow})
\hat{E}_L  - \frac{i\omega d}{\gamma}\hat{\tilde{F}}_{\downarrow},\nonumber\\
&&\hat{\rho}_{(11)\uparrow} = \frac{1}{\Gamma\hbar\omega}\left(\hat{j}^{\dag}_{\uparrow}\hat{E}_R 
+\hat{E}^{\dag}_R \hat{j}_{\uparrow}\right) + \frac{\hat{F}_{(11)\uparrow}}{\Gamma},\nonumber\\
&&\hat{\rho}_{(11)\downarrow} = \frac{1}{\Gamma\hbar\omega}\left(\hat{j}^{\dag}_{\downarrow}\hat{E}_L 
+\hat{E}^{\dag}_L \hat{j}_{\downarrow}\right) + \frac{\hat{F}_{(11)\downarrow}}{\Gamma}, 
\end{eqnarray}
where $\hat{\tilde{F}}_{\uparrow,\downarrow}e^{-i\omega t} = \hat{F}_{(10)\uparrow,\downarrow}$. 
We will assume that fluctuations in different valleys are not correlated:
\begin{equation}
\label{not-corr}
\langle \hat{\tilde{F}}^{\dag}_{\uparrow}\hat{\tilde{F}}_{\downarrow} \rangle = \langle \hat{F}_{(11)\uparrow}\hat{F}_{(11)\downarrow}\rangle = 0.
\end{equation}

For a single photon flux we can safely assume that the surface density of excited excitons is much lower than the maximum density of exciton states $N$  determined by the spectral bandwidth $\Delta\omega$ and the density of states. Then, taking into account the normalization condition
\begin{equation}
\hat{\rho}_{(00)\uparrow} + \hat{\rho}_{(11)\uparrow}
=\hat{\rho}_{(00)\downarrow} + \hat{\rho}_{(11)\downarrow}
=N,
\end{equation}
we can replace $\hat{\rho}_{(00)\uparrow} - \hat{\rho}_{(11)\uparrow} \rightarrow N$ and $\hat{\rho}_{(00)\downarrow} - \hat{\rho}_{(11)\downarrow} \rightarrow N$ in Eqs.~(\ref{stat}), which gives
\begin{equation}
\label{curr}
\hat{j}_{\uparrow} = \frac{\omega^2_{c\perp}\hat{E}_R}{4\pi\gamma}
 - \frac{i\omega d^*}{\gamma}\hat{\tilde{F}}_{\uparrow}, \, 
\hat{j}_{\downarrow} = \frac{\omega^2_{c\perp}\hat{E}_L}{4\pi\gamma}
 - \frac{i\omega d}{\gamma}\hat{\tilde{F}}_{\downarrow},
\end{equation}
where $\omega_{c\perp}^2 = \displaystyle \frac{4\pi\omega|d|^2 N}{\hbar}$ is the surface cooperative frequency squared which has the dimension of [cm/sec$^2$].

Equations (\ref{curr}) and the last two Eqs.~(\ref{stat}) give a complete description of the response of electron-hole or exciton states in a MX$_2$ monolayer to the field (\ref{field}). In particular, the polarization of the surface optical current excited in the monolayer and of the radiation emitted by the current can be found by calculating the Stokes parameters:
\begin{equation}
\label{stokes}
s_x = \langle\hat{j}^{\dag}_x\hat{j}_x \rangle,
s_y = \langle\hat{j}^{\dag}_y\hat{j}_y \rangle,
\end{equation}
where
$$
\hat{j}_x = \frac{1}{\sqrt{2}}(\hat{j}_{\uparrow} + \hat{j}_{\downarrow}), \;
\hat{j}_y = \frac{i}{\sqrt{2}}(\hat{j}_{\uparrow} - \hat{j}_{\downarrow}),
$$
and the brackets $\langle...\rangle$ mean averaging over the statistics of the reservoir and over the quantum state. 


\subsection{Self-consistent optical field in the monolayer}

Next, we need to relate the unknown field Eq.~(\ref{field}) in a monolayer to an incident field. For simplicity we will consider a 1D propagation problem, corresponding for example to a wide enough beam. If the monolayer is located in the z = 0 plane on a substrate with dielectric constant $\varepsilon = n^2$ for $z < 0$, the electromagnetic fields incident on the monolayer from both directions can be written as
\\

(i) Pump:
\begin{eqnarray}
\label{pump}
\hat{\textbf{E}}_i(z<0) & = & E_{\varepsilon}\left(
\textbf{e}_+e^{ikz-i\omega t}\hat{c}_R+\textbf{e}_-e^{-ikz+i\omega t}\hat{c}^{\dag}_R \right. \nonumber\\
&+&\left. \textbf{e}_-e^{ikz-i\omega t}\hat{c}_L+\textbf{e}_+e^{-ikz+i\omega t}\hat{c}^{\dag}_L
\right)
\end{eqnarray}

(ii) Vacuum noise:
\begin{eqnarray}
\label{vac}
\hat{\textbf{E}}_i(z>0) & = & E_{0}\left(
\textbf{e}_+e^{-ikz-i\omega t}\hat{c}_{VR} + \textbf{e}_-e^{ikz+i\omega t}\hat{c}^{\dag}_{VR} \right. \nonumber\\
&+&\left. \textbf{e}_-e^{-ikz-i\omega t}\hat{c}_{VL}+\textbf{e}_+e^{ikz+i\omega t}\hat{c}^{\dag}_{VL}
\right)
\end{eqnarray}
Here $E_{\varepsilon} = \sqrt{2\pi\hbar\omega/\varepsilon}$ and $E_0 = \sqrt{2\pi\hbar\omega}$ are the normalization amplitudes. 
In a more general and realistic case, the incident field is not exactly monochromatic but occupies a narrow frequency band $\Delta\omega\ll\omega $. In this case the creation and annihilation operators are defined for slowly varying amplitudes of the Heisenberg operators (see also \cite{mikhail,vdovin}).   Note that the operators $\hat{c}_R$ and $\hat{c}_{VR}$, and other similar combinations (same for $\hat{c}_L, \hat{c}_{VL}$) correspond to the field modes with different wave numbers: $k=|k|$ and $k=-|k|$. Therefore they always commute.

Normalization amplitudes in Eqs.~(\ref{pump},\ref{vac}) correspond to a unit quantization volume $V=1$, i.e.~the dyadics  $\hat{c}^{\dag}_R\hat{c}_R$ and $ \hat{c}^{\dag}_L\hat{c}_L$  are the operators of the photon density (see \cite{mikhail,vdovin,fain}), and the quantities like $\langle \hat{c}^{\dag}_R\hat{c}_R \rangle \equiv \langle \Psi_0 |\hat{c}^{\dag}_R\hat{c}_R| \Psi_0 \rangle$ and $\langle \hat{c}^{\dag}_L\hat{c}_L \rangle = \langle \Psi_0 |\hat{c}^{\dag}_L\hat{c}_L| \Psi_0 \rangle$ have the dimension of [cm$^{-3}$], where $\Psi_0$ is the initial state function present in the Heisenberg picture, which is normalized as $\langle\Psi_0|\Psi_0\rangle=1$ . 

When calculating relative average values, it is sufficient to take the field as monochromatic and use the explicit initial state $\Psi_0$. For example, if the initial field is in the linearly polarized single photon state equivalent to a single-photon Bell state in the circularly polarized basis, $\Psi_{0}  = \Psi_{\pm} $,
\begin{equation}
\label{psipm}
\Psi_{\pm} = \frac{|1_R\rangle|0_L\rangle\pm|0_R\rangle |1_L\rangle}{\sqrt{2}},
\end{equation}
the averages have the following properties:
\begin{eqnarray}
\label{init-bell}
&&\langle\hat{c}^{\dag}_R\hat{c}_R\rangle
=\langle\hat{c}^{\dag}_L\hat{c}_L\rangle =\pm\langle\hat{c}^{\dag}_R\hat{c}_L\rangle, \;
\langle\hat{c}^{\dag}_R\hat{c}_L\rangle = \langle\hat{c}^{\dag}_L\hat{c}_R\rangle; \nonumber\\
&&\langle\hat{c}^{\dag}_R\hat{c}_R\hat{c}^{\dag}_L\hat{c}_L\rangle = 0. 
\end{eqnarray}

If the incident field corresponds to the other Bell state, e.g.
\begin{equation}
\label{bell3}
\Phi_{\pm} = \frac{|0_R\rangle|0_L\rangle\pm|1_R\rangle|1_L\rangle}{\sqrt{2}},
\end{equation}
or if it is a state with the same average energy $\hbar \omega$ which is not entangled in the circularly polarized basis, e.g.
\begin{equation}
\label{not-ent}
\Psi_D = \frac{|0_R\rangle + e^{i\phi}|1_R\rangle}{\sqrt{2}}
\times
\frac{|0_L\rangle + e^{i\psi}|1_L\rangle}{\sqrt{2}},
\end{equation}
we obtain instead of Eqs.~(\ref{init-bell}) that 
\begin{eqnarray}
\label{init-bell2}
&&\langle\hat{c}^{\dag}_R\hat{c}_R\rangle
=\langle\hat{c}^{\dag}_L\hat{c}_L\rangle, \;  \langle\hat{c}^{\dag}_R\hat{c}_L\rangle = \langle\hat{c}^{\dag}_L\hat{c}_R\rangle = 0; \nonumber\\
&&\langle\hat{c}^{\dag}_R\hat{c}_R\hat{c}^{\dag}_L\hat{c}_L\rangle = \langle\hat{c}^{\dag}_R\hat{c}_R\rangle \langle\hat{c}^{\dag}_L\hat{c}_L\rangle. 
\end{eqnarray}

Whenever it is important to include Langevin noise terms, one has to take into account the finite spectral bandwidth of radiation $\Delta \omega \ll \omega$. In this case the properties of commutators and correlators are determined taking into account the density of photon states.  For example, for a paraxial beam with the aperture cross section $S_{\perp}$ we obtain following \cite{mikhail,vdovin} that
\begin{equation}
\label{parax}
\left[ \hat{c}_{\sigma\omega}\hat{c}^{\dag}_{\sigma\omega'} \right] = \frac{n}{2\pi cS_{\perp}}\delta(\omega-\omega').
\end{equation}
where $\hat{c}_{\sigma} = \int_{\Delta\omega}\hat{c}_{\sigma\omega}e^{-i\omega t}d\omega$, $n$ is the refractive index,  and the subscript $\sigma$ denotes the polarization of a normal mode of the field. For correlators of spectral components of the field satisfying $\langle \hat{\bf E} \rangle = 0$ we have
\begin{equation}
\label{delta}
S_{\perp}\frac{c}{n}\langle \hat{c}^{\dag}_{\sigma\omega}\hat{c}_{\sigma\omega'} \rangle
=\frac{N_{\sigma\omega}}{2\pi}\delta(\omega-\omega'),
\end{equation}
where the dimensionless quantity $N_{\sigma\omega}/2\pi$ determines the power $P_{\sigma\omega} = \hbar\omega(N_{\sigma\omega}/2\pi)$ incident on the area $S_{\perp}$ per unit frequency interval per unit time for a given polarization, so $P_{\sigma\omega}$ has a dimension of energy; for the vacuum field $N_{\sigma\omega} = 0$. 
Here we won't present cumbersome expressions for the state functions of multimode fields which are reduced to states (\ref{psipm}), (\ref{bell3}), or (\ref{not-ent}). We will only give some correlators needed for the subsequent derivation. In particular, conditions (\ref{init-bell}) and (\ref{init-bell2}) correspond to 
\begin{eqnarray}
\label{init-bell3}
&&\langle\hat{c}^{\dag}_R\hat{c}_R\rangle
=\langle\hat{c}^{\dag}_L\hat{c}_L\rangle \; \Rightarrow N_{R\omega} = N_{L\omega} \nonumber \\
&& \langle\hat{c}^{\dag}_R\hat{c}_L\rangle = \langle\hat{c}^{\dag}_L\hat{c}_R\rangle = \pm \langle\hat{c}^{\dag}_R\hat{c}_R\rangle \; \Rightarrow \langle\hat{c}^{\dag}_{R\omega}\hat{c}_{L\omega'}\rangle = \langle\hat{c}^{\dag}_{L\omega}\hat{c}_{R\omega'} \rangle = \pm \langle\hat{c}^{\dag}_{R\omega}\hat{c}_{R\omega'} \rangle  \\
&&\langle\hat{c}^{\dag}_R\hat{c}_L\rangle = \langle\hat{c}^{\dag}_L\hat{c}_R\rangle = 0 \; \Rightarrow \langle\hat{c}^{\dag}_{R\omega}\hat{c}_{L\omega'}\rangle = \langle\hat{c}^{\dag}_{L\omega}\hat{c}_{R\omega'} \rangle = 0 . \nonumber
\end{eqnarray}

Note that since a duration of a single-photon ''pulse'' is of the order of $\Delta\omega^{-1}$, the maximum power of a single photon illumination regime within the spectral bandwidth $\Delta\omega$ of of the order of $\hbar\omega\Delta\omega$ ; for a higher power individual photons overlap. Here we don't consider the possibility of using ''multiphoton'' fluxes of quantum correlated fields.

After performing a standard calculation of the operators of the reflected and transmitted fields, we obtain the following expression for the field operators $\hat{E}_{R,L}$ used in Eq.~(\ref{field}):
\begin{equation}
\label{ERL}
\hat{E}_{R,L} = \frac{2E_0}{n+1}\left(\hat{c}_{R,L}+\hat{c}_{VR,VL}\right) - \frac{4\pi \hat{j}_{\downarrow,\uparrow}}{(n+1)c} .
\end{equation}
The last term on the right-hand side of Eq.~(\ref{ERL}) gives rise to the the collective superradiant relaxation of the excitonic ensemble with the rate $\Omega_{SR} = \displaystyle \frac{\omega_{c\perp}^2}{(n+1) c}$ \cite{kono2014}. Taking into account Eqs.~(\ref{curr}) it is easy to find that this term can be neglected provided $\gamma \gg \Omega_{SR}$. In this case one can neglect radiative corrections (back reaction) when calculating the current excited in the monolayer. 


\subsection{The effect of radiative corrections}

To study the effect of radiative corrections on the excited state populations $\langle\hat{\rho}_{11}\rangle$ it is convenient to introduce a space-time spectrum of noise, 
\begin{equation}
\label{noise}
\hat{F}_{mn} = \int_{\infty}\hat{F}_{q,v,;mn}e^{i {\bf q} {\bf r}_{\perp}-ivt}dvd^2q, \hat{F}_{q,v;mn} = \hat{F}^{\dag}_{-q,-v;nm};
\end{equation}
we can obtain from (\ref{corr}) the correlators for the spectral components of the noise operator:
\begin{eqnarray}
\label{spectrum}
&&\langle \hat{F}_{q,v;mn}^{\dag}\hat{F}_{q',v';mn}\rangle
=
\frac{1}{4\pi^3}\left( \gamma_{mn}\langle\hat{\rho}_{mm}\rangle + \frac{1}{2}\langle\hat{R}_{mm}\rangle\right) \delta(v-v')\delta(\textbf{q}- \textbf{q'})\nonumber\\
&&\langle \hat{F}_{q,v;mn}\hat{F}^{\dag}_{q',v';mn}\rangle =
\frac{1}{4\pi^3}\left( \gamma_{mn}\langle\hat{\rho}_{nn}\rangle + \frac{1}{2}\langle\hat{R}_{nn}\rangle\right) \delta(v-v')\delta(\textbf{q}- \textbf{q'}).
\end{eqnarray}
Note that the harmonics of the current with wave vectors $q > \omega/c$ cannot excite propagating modes and therefore cannot give rise to radiative losses. Therefore we can put $q,q' \leq \omega/c$ in Eq.~(\ref{spectrum}) and, using Eqs.~(\ref{stat},\ref{curr},\ref{ERL},\ref{spectrum}), after lengthy but straightforward calculations, obtain an estimate of the radiative relaxation rate of populations induced by noise: 
\begin{equation}
\Omega_{rad} \simeq \frac{2\omega |d|^2}{c\hbar} \int_{0}^{\omega/c} q\,dq.
\end{equation}
As expected, this rate corresponds to the inverse spontaneous emission time. 
 
 To summarize, when calculating the excited state populations $\langle\hat{\rho}_{11}\rangle$ the radiative corrections can be neglected if $2 \Omega_{SR}, \Omega_{rad} \ll \Gamma$. Taking into account the back reaction effects will make the derivation more complicated, and the result will amount to renormalization of the relaxation rates in the final expressions:
$$
\gamma \rightarrow \gamma + \Omega_{SR}, \Gamma \rightarrow \Gamma + 2\Omega_{SR} + \Omega_{rad}.
$$
To keep the derivation more streamlined, we will ignore the radiative corrections and assume that 
\begin{equation}
\label{ERL2}
\hat{E}_{R,L} = \frac{2E_0}{n+1}(\hat{c}_{R,L}+\hat{c}_{VR,VL}).
\end{equation}
 
 
 \subsection{The polarization of the photoluminescence from excitons excited by a single-photon field}

The initial (Heisenberg) state of our system can be represented as
\begin{eqnarray}
\label{init2}
&&|\Psi_0\rangle = | \Psi_{\uparrow\downarrow} \rangle |\Psi_{V} \rangle | \Psi_{(R+L)} \rangle, \;
|\Psi_{\uparrow\downarrow}\rangle = |0\uparrow\rangle|0\downarrow\rangle,\nonumber\\
&&|\Psi_V\rangle = |0_{VR}\rangle|0_{VL}\rangle.
\end{eqnarray}
First let us neglect the contribution of the Langevin sources. Consider an incident field in the Bell state
\begin{equation}
\label{bell2}
|\Psi_{R+L}\rangle = \Psi_{\pm},
\end{equation}
which corresponds to a single-photon linearly polarized field $|1_x\rangle|0_y\rangle$ or $|0_x\rangle|1_y\rangle$ respectively (see Appendix B). If we substitute expression (\ref{ERL2}) in Eqs.~(\ref{curr}) and (\ref{stokes}), then after performing the averaging in Eqs.~(\ref{stokes}) all dyadics that include the vacuum field operators will become zero. A non-zero contribution comes only from averaging the dyadics $\hat{c}^{\dag}_R\hat{c}_R$, $\hat{c}^{\dag}_L\hat{c}_L$, $\hat{c}^{\dag}_R\hat{c}_L$, and $\hat{c}^{\dag}_L\hat{c}_R$:
\begin{eqnarray}
\label{dyad}
&&\langle \Psi_0|\hat{c}^{\dag}_R\hat{c}_R|\Psi_0\rangle = \langle \Psi_0|\hat{c}^{\dag}_L\hat{c}_L|\Psi_0\rangle = \frac{1}{2},\nonumber\\
&&\langle \Psi_0|\hat{c}^{\dag}_R\hat{c}_L|\Psi_0\rangle = \langle \Psi_0|\hat{c}^{\dag}_L\hat{c}_R|\Psi_0\rangle =\pm \frac{1}{2}.
\end{eqnarray}

As  a result we obtain $s_x\ne 0, s_y = 0$ for $|\Psi_{R+L}\rangle = \Psi_+$ and $s_y\ne 0, s_x = 0$ for $|\Psi_{R+L}\rangle = \Psi_-$. In other words, the reemitted photons will have linear polarization as in the cavity geometry described in Sec.~II. At the same time, the reemitted photons will be unpolarized, $s_x = s_y$, if the incident field corresponds to other Bell states of a single-photon energy, Eq.~(\ref{bell3}), or if it is a single-photon state which is not entangled, Eq.~(\ref{not-ent})

So what is the role of the exciton entanglement here? The fact that a linearly polarized 	radiation remains linearly polarized after interacting with a system of equal numbers of ''left'' and ''right'' rotators (such  as K and K' valleys in MoS$_2$) does not by itself constitute the evidence that left and right subsystems are entangled. Indeed, the same result can be obtained for a classical field \cite{jones2013} which cannot entangle the non-interacting quantum systems; see Appendix A. However, for a quantum incident field, e.g.~for a single-photon field the conservation of the linear polarization in the reemitted radiation is possible only if left and right excitons are entangled. Therefore, in this case the degree of the linear polarization of the reemitted field can be used to verify the exciton valley entanglement.

In order to get an explicit expression for the noise-induced depolarization of the reemitted field, we need to sum over all modes in a spectral bandwidth $\Delta \omega$. If we take the Langevin noise terms into account in Eqs.~(\ref{curr}) and use Eqs.~(\ref{parax}) and (\ref{delta}), the Stokes parameters for the initial state $|\Psi_{R+L} \rangle = \Psi_+$  are
\begin{eqnarray}
&&s_x = 2\left( \frac{\omega_{c\perp}^2E_0}{2\pi\gamma(n+1)} \right)^2 \frac{n(N_{R\omega} + N_{L\omega})\Delta \omega}{2\pi c S_{\perp}} + \Lambda, \; s_y = \Lambda,\\
&&\Lambda = \frac{\omega^2|d|^2}{\gamma^2}\langle \hat{\tilde{F}}^{\dag}_{\uparrow}\hat{\tilde{F}}_{\uparrow}
+ \hat{\tilde{F}}^{\dag}_{\downarrow}\hat{\tilde{F}}_{\downarrow} \rangle.
\end{eqnarray}

When calculating the value of $\Lambda$ we take into account only the spatial harmonics of the Langevin noise within the spectral bandwidth $\Delta \omega$ that correspond to a paraxial beam of the aperture $S_{\perp}$. We will also assume that the relaxation of coherence is much faster that the recombination rate, $\gamma\gg\Gamma$, which is typically the case in semiconductors. Using also Eq.~(\ref{spectrum}), we arrive at
\begin{eqnarray}
\label{lambda}
&&\Lambda = \omega^2|d|^2\Delta\omega
\frac{\langle\Psi_0|\hat{\rho}_{(11)\uparrow}|\Psi_0\rangle + \langle\Psi_0|\hat{\rho}_{(11)\downarrow}|\Psi_0\rangle}
{2\pi\gamma S_\perp}, \nonumber \\
&& \frac{s_y}{s_x} = \frac{\alpha}{\alpha + 1}, \\
&&\alpha = \frac{\langle\Psi_0|\hat{\rho}_{(11)\uparrow}|\Psi_0\rangle + \langle\Psi_0|\hat{\rho}_{(11)\downarrow}|\Psi_0\rangle}{2N(N_{R\omega}+N_{L\omega})}\times\frac{c\gamma}{\omega^2_{c\perp}}
\times\frac{(n+1)^2}{2n}.\nonumber 
\end{eqnarray}

In order to find the fraction of the unpolarized field $\alpha$ we need to calculate the operators $\hat{\rho}_{(11)\uparrow}$ and $\hat{\rho}_{(11)\downarrow}$ in Eqs.~(\ref{lambda}). Substituting (\ref{curr}) and (\ref{ERL2})  into the last two Eqs.~(\ref{stat}) it is easy to find that the resulting noise-dependent terms are linear with respect to the Langevin noise operators, so that they vanish after averaging over the reservoir. As a result, we obtain
\begin{eqnarray}
\label{den-corr}
&&\hat{\rho}_{(11)\uparrow} = \chi (\hat{c}^\dag_R + \hat{c}^\dag_{VR})(\hat{c}_R + \hat{c}_{VR}),\nonumber\\
&&\hat{\rho}_{(11)\downarrow} = \chi (\hat{c}^\dag_L + \hat{c}^\dag_{VL})(\hat{c}_L + \hat{c}_{VL}),
\end{eqnarray}
where $\chi = \displaystyle \frac{\omega^2_{c\perp}|E_0|^2}{\pi\gamma\Gamma(n+1)^2\hbar\omega}$.
After summing over the modes within the bandwidth $\Delta\omega$ and  taking into account Eqs.~(\ref{parax}),(\ref{delta}), we get
\begin{equation}
\label{11}
\langle\Psi_0|\hat{\rho}_{(11)\uparrow}|\Psi_0\rangle
=\langle\Psi_0|\hat{\rho}_{(11)\downarrow}|\Psi_0\rangle
=\frac{n}{(n+1)^2}\frac{\omega_{c\perp}^2}{\gamma c \Gamma S_\perp}\frac{N_{(R,L)\omega}\Delta\omega}{2\pi}.
\end{equation}
Substituting Eq.~(\ref{11}) into (\ref{lambda}) yields
\begin{equation}
\label{alpha}
\alpha = \frac{\Delta \omega}{2\pi \Gamma}\frac{1}{2NS_\perp},
\end{equation}
Where the ratio $N/\Delta \omega$ is determined by the exciton or electron-hole density of states. Equation (\ref{alpha}) has a simple physical interpretation: a fraction of the unpolarized field in the reemitted radiation is equal to the ratio of a number of photons incident on a monolayer during the lifetime of an exciton to the total number of exciton states within $\Delta \omega$. As is clear from Eq.~(\ref{alpha}), $\alpha$ scales as one divided by the number of exciton or electron-hole states within the bandwidth $\Gamma \sim 10^9-10^{10}$  s$^{-1}$, so assuming a standard 2D density of states $m/(\pi\hbar)$ for parabolic bands and the beam aperture $S_\perp > \lambda^2 \sim 10^{-8}$ cm$^{-2}$, $\alpha$ is smaller than $10^{-2}$.   


\subsection{Correlation properties of photoexcited excitons}

It is easy to show that in the case of the initial Bell state of the field given by Eq.~(\ref{psipm}) the correlator for the density operators (\ref{den-corr}) obeys the following property:
\begin{equation}
\label{prop}
\langle\Psi_0|\hat{\rho}_{(11)\uparrow}\hat{\rho}_{(11)\downarrow}|\Psi_0\rangle = 0,
\end{equation}
which coincides with the corresponding relationship (\ref{pair}) for the solution to the Schroedinger equation for a cavity field. It is important that when Eqs.~(\ref{not-corr}) are satisfied, the condition (\ref{prop}) is not affected by the Langevin noise.
	
For unpolarized but still single-photon quantum fields (\ref{bell3}),(\ref{not-ent}) we obtain instead of (\ref{prop}) that
\begin{equation}
\langle\Psi_0|\hat{\rho}_{(11)\uparrow}\hat{\rho}_{(11)\downarrow}|\Psi_0\rangle = \langle\Psi_0|\hat{\rho}_{(11)\uparrow}|\Psi_0\rangle
\times\langle\Psi_0|\hat{\rho}_{(11)\downarrow}|\Psi_0\rangle,
\end{equation}
which is the same as for classical fields. Therefore, Eq.~(\ref{prop}) is a unique property of the valley-entangled exciton system created  by an incident quantum field in the initial state (\ref{bell2}). Excitons created by a classical field or by quantum fields in the initial states (\ref{bell3},\ref{not-ent}) do not satisfy Eq. (\ref{prop}).


\section{Photocurrent in a valley-entangled electron-hole system}

Quantum correlations between photoexcited carriers should manifest themselves in the fluctuations of the photocurrent or photovoltage under the applied bias. A DC electric field applied along the monolayer can lead to separation of photoexcited electrons from holes while not affecting their valley index and any possible entanglement with respect to the valley degree of freedom; see Fig.~3. This will create a photocurrent or photovoltage depending on the way the detector is wired in an external circuit. Note that breaking the binding energy of excitons may require a very high electric field. In experiments  \cite{mak2014} the charge separation most likely originated from the metal-semiconductor contact regions with a high built-in electric field. 

\begin{figure}[htbp]
\includegraphics[width=0.5\textwidth]{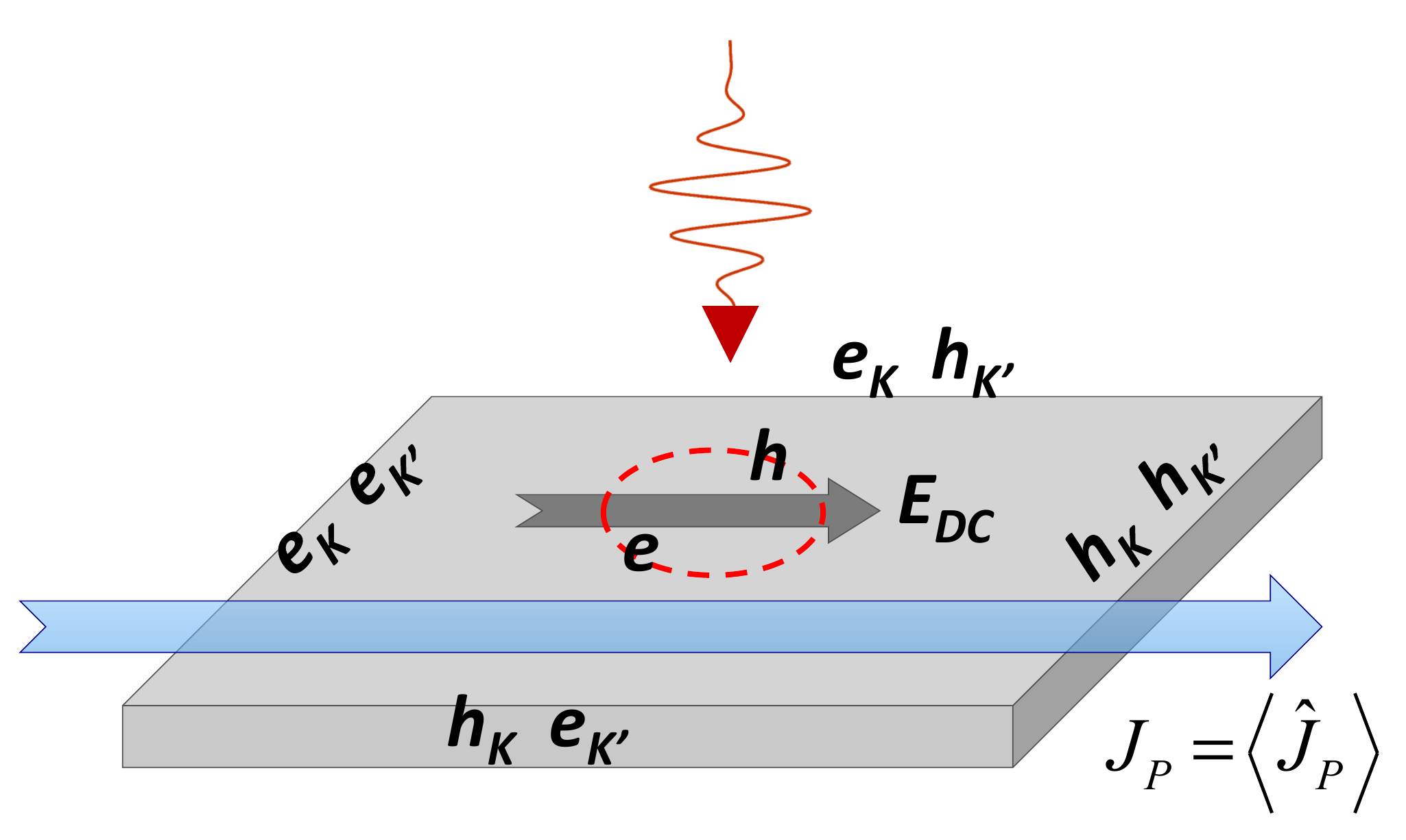}
\caption{(Color online) A sketch of photocurrent detection. Under illumination with a linearly polarized single-photon field and in the presence of an in-plane DC electric field ${\bf E_{DC}}$ there is charge separation in the direction along ${\bf E_{DC}}$ which gives rise to a photocurrent $J_P$ if a monolayer is contacted and included in a proper external circuit. There is no photocurrent or photovoltage in the transverse direction. }
\end{figure}


A lot of attention has been recently devoted to the valley Hall effect which generates photovoltage in the transverse direction under a circularly polarized excitation \cite{xiao2012,mak2014}.  However, for a linearly polarized excitation there will be zero Hall voltage with equal amounts of positive and negative charges accumulated on the sides as sketched in Fig.~3. Here we consider the longitudinal photocurrent along the direction of the applied DC field.  Specifically, the current detector will measure the quantity $J_P = \langle\Psi_0|\hat{J}_P|\Psi_0\rangle$, where the averaging is taken over the initial state (\ref{init2}) and $\hat{J}_P$ is the photocurrent operator:
\begin{equation}
\label{current}
\hat{J}_P = \eta (\hat{\rho}_{(11)\uparrow}+\hat{\rho}_{(11)\downarrow}),
\end{equation}
where $\eta$ is a certain coefficient. Obviously, equal amounts of photogenerated electrons and holes from K and K' valleys will contribute to the signal. When characterizing quantum correlation properties of carriers, the quantity of interest is not the current itself but current fluctuations defined as 
\begin{equation}
\label{cur-fluct}
\delta J_P = \sqrt{\langle \hat{J}_P^2 \rangle - \langle \hat{J}_P \rangle^2}.
\end{equation}

Using the initial quantum state defined by Eq.~(\ref{init2}), we compare the current fluctuations due to the following three states of an incident field $\Psi_{R+L}$ that have the same average energy:

(i) Unpolarized field:
\begin{equation}
\label{not-ent2}
\Psi_{R+L} = \Psi_D = \frac{|0_R\rangle + e^{i\phi}|1_R\rangle}{\sqrt{2}}
\times
\frac{|0_L\rangle + e^{i\psi}|1_L\rangle}{\sqrt{2}}.
\end{equation}

(ii) Circularly polarized field:
\begin{equation}
\label{circ}
\Psi_{R+L} = \Psi_R = |1_R\rangle |0_L\rangle. 
\end{equation}

(iii) Linearly polarized field that can be represented as an entangled state of two circular polarizations (see Appendix B): 
\begin{equation}
\label{psiplus}
\Psi_{R+L} = \Psi_+ = \frac{|1_R\rangle|0_L\rangle + |0_R\rangle |1_L\rangle}{\sqrt{2}}. 
\end{equation}

To calculate the photocurrent fluctuations, we need to evaluate the averages of both  the excited-state density operators in Eqs.~(\ref{den-corr}) and their products. The first of them gives 
\begin{equation}
\label{up1}
\hat{\rho}_{(11)\uparrow} = \chi (\hat{c}^\dag_R\hat{c}_R + \hat{c}^\dag_{VR}\hat{c}_R + \hat{c}^\dag_R\hat{c}_{VR} + \hat{c}^\dag_{VR}\hat{c}_{VR});
\end{equation} 
\begin{eqnarray}
\label{up2}
&&\hat{\rho}_{(11)\uparrow} \hat{\rho}_{(11)\uparrow}
=\chi^2( \hat{c}^\dag_R\hat{c}_R\hat{c}^\dag_R\hat{c}_R + \hat{c}^\dag_{VR}\hat{c}^\dag_{VR}\hat{c}_R\hat{c}_R\nonumber\\
&& + \hat{c}^\dag_R\hat{c}^\dag_R\hat{c}_{VR}\hat{c}_{VR} + \hat{c}^\dag_{VR}\hat{c}_{VR}\hat{c}^\dag_{VR}\hat{c}_{VR}
+\hat{c}^\dag_{VR}\hat{c}_R\hat{c}^\dag_R\hat{c}_R\nonumber\\
&&+\hat{c}^\dag_R\hat{c}^\dag_R\hat{c}_R\hat{c}_{VR}+\hat{c}^\dag_R\hat{c}_R\hat{c}^\dag_{VR}\hat{c}_{VR}
+\hat{c}^\dag_R\hat{c}_R\hat{c}_R\hat{c}^\dag_{VR}\nonumber\\
&&+\hat{c}_{VR}\hat{c}^\dag_{VR}\hat{c}^\dag_R\hat{c}_R + \hat{c}^\dag_{VR}\hat{c}^\dag_{VR}\hat{c}_{VR}\hat{c}_R
+\hat{c}^\dag_R\hat{c}_R\hat{c}^\dag_R\hat{c}_{VR}\nonumber\\
&&+\hat{c}^\dag_{VR}\hat{c}_{VR}\hat{c}_R\hat{c}^\dag_R +\hat{c}^\dag_{VR}\hat{c}_{VR}\hat{c}_{VR}\hat{c}^\dag_R
+\hat{c}^\dag_R\hat{c}_R\hat{c}^\dag_{VR}\hat{c}_{VR}\nonumber\\
&&+\hat{c}^\dag_{VR}\hat{c}^\dag_{VR}\hat{c}_{VR}\hat{c}_R+\hat{c}_{VR}\hat{c}^\dag_{VR}\hat{c}_{VR}\hat{c}^\dag_R).
\end{eqnarray}

The same result can be obtained for the population of the ''$\downarrow$'' states after replacing $\uparrow$ with $ \downarrow$ and $R$ with $L$.  After averaging (\ref{up1}) and (\ref{up2}) over the initial quantum state Eq.~(\ref{init2}) and taking into account the commutation relations we obtain:
\begin{equation}
\label{aver} 
\langle \hat{\rho}_{(11)\uparrow} \rangle
=\chi\langle\Psi_{R+L}| \hat{c}^\dag_R\hat{c}_R |\Psi_{R+L}\rangle, \; \langle \hat{\rho}_{(11)\downarrow} \rangle
=\chi\langle\Psi_{R+L}| \hat{c}^\dag_L\hat{c}_L |\Psi_{R+L}\rangle,
\end{equation}
\begin{eqnarray}
\label{fluct}
&&\langle \hat{\rho}_{(11)\uparrow}\hat{\rho}_{(11)\uparrow} \rangle = \chi^2\langle\Psi_{R+L}| \hat{c}^\dag_R\hat{c}_R |\Psi_{R+L}\rangle \left(1 +  \langle 0_{VR}|\hat{c}_{VR}\hat{c}^\dag_{VR}|0_{VR}\rangle \right) \nonumber\\
&&\langle \hat{\rho}_{(11)\downarrow}\hat{\rho}_{(11)\downarrow} \rangle = \chi^2\langle\Psi_{R+L}| \hat{c}^\dag_L\hat{c}_L |\Psi_{R+L}\rangle \left(1 +  \langle 0_{VL}|\hat{c}_{VL}\hat{c}^\dag_{VL}|0_{VL}\rangle \right).
\end{eqnarray}
As is clear from Eqs.~(\ref{fluct}), vacuum fluctuations of the field amplify the fluctuations of the photocurrent. In the absence of an incident field $J_P = \delta J_P = 0$ as expected. It also follows from from Eqs.~(\ref{aver}), (\ref{fluct}) that the following relations are true for all three states of the field:
\begin{equation}
\label{aver2} 
\langle \hat{\rho}_{(11)\uparrow} \rangle = \langle \hat{\rho}_{(11)\downarrow} \rangle = \chi, \; 
\langle \hat{\rho}_{(11)\uparrow}\hat{\rho}_{(11)\uparrow} \rangle = \langle \hat{\rho}_{(11)\downarrow}\hat{\rho}_{(11)\downarrow} \rangle = 2 \chi^2. 
\end{equation}

Next, we calculate the averages for ''mixed'' products $\hat{\rho}_{(11)\uparrow}\hat{\rho}_{(11)\downarrow}$ and $\hat{\rho}_{(11)\downarrow}\hat{\rho}_{(11)\uparrow}$:
\begin{equation}
\label{mixed} 
\langle \hat{\rho}_{(11)\uparrow}\hat{\rho}_{(11)\downarrow} \rangle = \langle \hat{\rho}_{(11)\downarrow}\hat{\rho}_{(11)\uparrow} \rangle 
= 2\chi^2\langle\Psi_{R+L}| \hat{c}^\dag_R\hat{c}_R \hat{c}^\dag_L\hat{c}_L|\Psi_{R+L}\rangle.
\end{equation}
Proceeding in the same way as above, for an unpolarized field (\ref{not-ent2}) we obtain 
 $\langle \hat{\rho}_{(11)\uparrow}\hat{\rho}_{(11)\downarrow} \rangle = \langle \hat{\rho}_{(11)\uparrow} \rangle \times \langle \hat{\rho}_{(11)\downarrow} \rangle$, whereas for polarized fields (\ref{circ}) and (\ref{psiplus}) $\langle \hat{\rho}_{(11)\uparrow}\hat{\rho}_{(11)\downarrow} \rangle = \langle \hat{\rho}_{(11)\downarrow}\hat{\rho}_{(11)\uparrow} \rangle = 0$. This result is trivial for a circularly polarized field (\ref{circ}) since the latter does not interact with ''$\downarrow$'' excitons. For a linearly polarized field (\ref{psiplus}) this result is not obvious and is entirely due to the entanglement of ''$\downarrow$'' and ''$\uparrow$''  excitons. 
 
 Using Eqs.~(\ref{aver2}) and (\ref{mixed}) to calculate the current (\ref{current}) and its fluctuations (\ref{cur-fluct}) we obtain $\delta J_P = \sqrt{2} J_P$ for an illumination with an unpolarized field and  $\delta J_P = J_P$ in the case of a linearly polarized incident field. This reduction of fluctuations by $\sqrt{2}$ is the direct consequence of the interference within the valley-entangled state of photoexcited carriers. 

In conclusion, we have shown that excitons in K and K' valleys transition metal dichalcogenide monolayers can be efficiently entangled by interacting with linearly polarized single photons. Valley entanglement leads to linear polarization of reemitted photons and squeezing of the photocurrent fluctuations. 

\section*{Acknowledgments}
We are grateful to Maria Erukhimova, Yevgeni Radeonychev, and Joseph Tokman for helpful discussions. This work has been supported by NSF Grants OISE-0968405 and EEC-0540832, and by the Air Force Office for Scientific Research. M. D. Tokman acknowledges support by the Russian Foundation for Basic Research Grants No. 13-02-00376 and No. 14-22-02034.

\section{Appendix} 

\subsection{Conditions for entanglement with classical fields}

Consider two quantum systems described by generalized coordinates $q_1$ and $q_2$. Their wave function $\Psi(q_1, q_2,t)$ obeys Schroedinger's equation
\begin{equation}
\label{a1}
i \hbar \dot{\Psi} = \hat{H} \Psi.
\end{equation} 
Assume that these two systems are not coupled with each other directly but interact with a classical electromagnetic field described by a classical variable $u(t)$.  In this case the Hamiltonian in Eq.~(\ref{a1}) can be written as $\hat{H} = \hat{H}_1(q_1,u(t)) + \hat{H}_2(q_2,u(t))$ where the operators $\hat{H}_1$ and  $\hat{H}_2$ act only on the functions of variables $q_1$ and $q_2$, respectively. Then, substituting $\Psi = \psi_1(q_1,t) \psi_2(q_2,t)$ into Eq.~(\ref{a1}), we obtain 
\begin{equation}
\label{a2}
\psi_2 \left( i\hbar \dot{\psi}_1 - \hat{H}_1 \psi_1 \right) + \psi_1 \left( i\hbar \dot{\psi}_2 - \hat{H}_2 \psi_2 \right) = 0.
\end{equation} 
If the quantum systems were not entangled at the initial moment of time $t = 0$, i.e. $\Psi(q_1, q_2,0) = \psi_1(q_1,0) \times \psi_2(q_2,0)$, Eq.~(\ref{a2}) splits into two independent equations for each system:
\begin{equation}
\label{a3}
i \hbar \dot{\psi}_1 = \hat{H}_1 \psi_1, \; i \hbar \dot{\psi}_2 = \hat{H}_1 \psi_2,
\end{equation} 
and the solution will remain in the form of the direct product $\Psi(q_1, q_2,t) = \psi_1(q_1,t) \times \psi_2(q_2,t)$, corresponding to unentangled systems. 

Entanglement may appear if the classical field gives rise to the interaction Hamiltonian $\hat{V}(u(t), q_1, g_2)$ that directly couples the two systems, for example: 
\begin{equation} 
\hat{H} = \hat{H}_1(q_1) + \hat{H}_2(q_2) + \chi u(t) q_1 q_2, 
\end{equation}
where $\chi$ is a coupling constant. This particular example corresponds to the Hamiltonian describing generation of entangled photons in a medium with a second order nonlinearity as a result of parametric frequency conversion \cite{scully}. 

\subsection{Relationship between $XY$ and $RL$ basis states for photons}

Consider the quantum field in vacuum in a quantization volume $V$:
\begin{equation}
\label{b1}
\hat{\bf E} = E_0 \left( {\bf x_0} e^{ikz - i\omega t} \hat{c}_x + {\bf x_0} e^{-ikz + i\omega t} \hat{c}_x^{\dag} + {\bf y_0} e^{ikz - i\omega t} \hat{c}_y + {\bf y_0} e^{-ikz + i\omega t} \hat{c}_y^{\dag} \right),
\end{equation} 
where $ {\bf x_0,y_0}$ are unit vectors along $x,y$ coordinate axes, 
\begin{equation}
\label{b2}
\int_V \sin kz \, d^3r = \int_V \cos kz \, d^3r = \int_V \sin 2kz \, d^3r = \int_V \cos 2kz \, d^3r = 0,
\end{equation} 
$$ 
E_0 = \sqrt{\frac{2 \pi \hbar \omega}{V}}.
$$
Let's expand the field given by Eq.~(\ref{b1}) in terms of circularly polarized orthogonal modes:
\begin{equation}
\label{b3}
\hat{\bf E} = E_0 \left( {\bf e_+} e^{ikz - i\omega t} \hat{c}_R + {\bf e_-} e^{-ikz + i\omega t} \hat{c}_R^{\dag} + {\bf e_-} e^{ikz - i\omega t} \hat{c}_L + {\bf e_+} e^{-ikz + i\omega t} \hat{c}_L^{\dag} \right),
\end{equation} 
where ${\bf e_{\pm}} = \frac{{\bf x_0} \pm i {\bf y_0}}{\sqrt{2}}$. Comparing Eqs.~(\ref{b1}) and (\ref{b3}) we obtain 
\begin{equation}
\label{b4}
\hat{c}_{x,y} = \frac{\hat{c}_R\pm \hat{c}_L}{\sqrt{2}}, \; \hat{c}_{R,L} = \frac{\hat{c}_x \mp \hat{c}_y}{\sqrt{2}}.
\end{equation} 

Next, we introduce the vacuum state $|0_{\Sigma} \rangle$; $\langle 0_{\Sigma} ||0_{\Sigma} \rangle = 1$. By definition of creation and annihilation operators, 
\begin{equation}
\label{b5}
|N_x\rangle |N_y \rangle = \frac{\left(\hat{c}_x^{\dag} \right)^{N_x} \left(\hat{c}_y^{\dag} \right)^{N_y} |0_{\Sigma}\rangle }{\sqrt{N_x! N_y!}}, \;  |N_R\rangle |N_L \rangle = \frac{\left(\hat{c}_R^{\dag} \right)^{N_R} \left(\hat{c}_L^{\dag} \right)^{N_L} |0_{\Sigma}\rangle}{\sqrt{N_R! N_L!}}.
\end{equation} 

Expanding the linearly polarized photon state 
\begin{equation}
\label{b6}
\Psi = |1_x\rangle |0_y\rangle
\end{equation}
in the basis of circularly polarized modes, we obtain: 
\begin{equation}
\label{b7}
\Psi = \sum_{N_R = 0, N_L = 0}^{\infty, \infty} A_{N_R,N_L} |N_R\rangle |N_L \rangle.
\end{equation}
Taking into account Eq.~(\ref{b4}), one can obtain from Eqs.~(\ref{b6}) and (\ref{b7}) that 
\begin{eqnarray}
A_{N_R,N_L}  = \langle N_R| \langle N_L||1_x \rangle |0_y \rangle = \langle 0_{\Sigma} | \frac{(\hat{c}_R)^{N_R} (\hat{c}_L)^{N_L} }{\sqrt{N_R! N_L!} } \hat{c}_x^{\dag} |0_{\Sigma} \rangle \nonumber \\
= \langle 0_{\Sigma} | \frac{(\hat{c}_R)^{N_R} (\hat{c}_L)^{N_L} \left( \hat{c}_R^{\dag} + \hat{c}_L^{\dag}\right) }{\sqrt{2N_R! N_L!} } |0_{\Sigma} \rangle = \left\{\begin{array}{cc}
0 & {\rm if} \; N_R + N_L \neq 1 \\
A_{1_R 0_L} = A_{0_R 1_L} = \frac{1}{\sqrt{2}} & {\rm if} \; N_R + N_L = 1.
\end{array}\right.
\label{b8}
\end{eqnarray}

As is clear from Eq.~(\ref{b8}), the product state $ |1_x\rangle |0_y\rangle$ in the basis of linearly polarized modes is one of the Bell states in the basis of circularly polarized modes: 
\begin{equation}
\label{b9}
\Psi_+ = \frac{|1_R\rangle |0_L\rangle + |0_R\rangle |1_L\rangle }{\sqrt{2}}. 
\end{equation}

The relationship $ |1_x\rangle |0_y\rangle = \Psi_+$ can be also obtained in a less formal way from average values of squares of cartesian components of the field. Indeed, for the field given by Eqs.~(\ref{b1}), (\ref{b2}), and (\ref{b3}) one can obtain the following expressions for the averages:
\begin{equation}
\label{b10}
\frac{1}{4 \pi} \left\langle \int_V \hat{\bf E} \hat{\bf E} \, d^3r \right \rangle = 2 \hbar \omega, \;  \frac{1}{4 \pi} \left\langle \int_V \hat{E}_x \hat{E}_x \, d^3r \right \rangle = \hbar \omega \left(1+\frac{1}{2} \right)  , \;   \frac{1}{4 \pi} \left\langle \int_V \hat{E}_y \hat{E}_y \, d^3r \right \rangle =  \frac{\hbar \omega}{2} .  
\end{equation}
Using the relationships
\begin{equation}
\label{b11}
\hat{E}_x = \frac{E_0}{\sqrt{2}} \left( e^{ikz - i \omega t} \hat{c}_R + e^{-ikz + i \omega t} \hat{c}_R^{\dag}  + e^{ikz - i \omega t} \hat{c}_L + e^{-ikz + i \omega t} \hat{c}_L^{\dag} \right),  
\end{equation}
\begin{equation}
\label{b12}
\hat{E}_y = \frac{E_0}{\sqrt{2}} \left( i e^{ikz - i \omega t} \hat{c}_R - i e^{-ikz + i \omega t} \hat{c}_R^{\dag}  - i e^{ikz - i \omega t} \hat{c}_L + i e^{-ikz + i \omega t} \hat{c}_L^{\dag} \right),  
\end{equation}
we can express the averages through creation and annihilation operators of circularly polarized modes: 
\begin{equation}
\label{b13}
\frac{1}{4 \pi} \left\langle \int_V \hat{\bf E} \hat{\bf E} \, d^3r \right \rangle =  \frac{\hbar \omega}{2} \left\langle \hat{c}_R \hat{c}_R^{\dag} + \hat{c}_R^{\dag} \hat{c}_R  +  \hat{c}_L \hat{c}_L^{\dag} + \hat{c}_L^{\dag} \hat{c}_L \right \rangle,   
\end{equation}
\begin{equation}
\label{b14}
\left\langle \int_V \hat{E}_x \hat{E}_x \, d^3r \right \rangle =  \frac{\hbar \omega}{4} \left\langle \hat{c}_R \hat{c}_R^{\dag} + \hat{c}_R^{\dag} \hat{c}_R  +  \hat{c}_L \hat{c}_L^{\dag} + \hat{c}_L^{\dag} \hat{c}_L + 2 \hat{c}_R \hat{c}_L^{\dag} + 2\hat{c}_R^{\dag} \hat{c}_L \right \rangle,   
\end{equation}
\begin{equation}
\label{b15}
\left\langle \int_V \hat{E}_y \hat{E}_y \, d^3r \right \rangle =  \frac{\hbar \omega}{4} \left\langle \hat{c}_R \hat{c}_R^{\dag} + \hat{c}_R^{\dag} \hat{c}_R  +  \hat{c}_L \hat{c}_L^{\dag} + \hat{c}_L^{\dag} \hat{c}_L - 2 \left( \hat{c}_R \hat{c}_L^{\dag} + 2\hat{c}_R^{\dag} \hat{c}_L \right) \right \rangle.    
\end{equation}
The first of Eqs.~(\ref{b10}) can be satisfied by any states of the type
\begin{equation}
\label{b16}
\Psi = A |1_R\rangle |0_L\rangle + B |0_R\rangle   |1_L\rangle ,  
\end{equation}
if 
\begin{equation}
\label{b17}
|A|^2 = |B|^2 = \frac{1}{2}.   
\end{equation}
Taking into account the last two equations in Eq.~(\ref{b10}), we obtain
$$
\left( |A|^2 + |B|^2 \right) + \frac{1}{2} \left( AB^* + A^*B \right) = \frac{3}{2}, 
$$
$$ 
\left( |A|^2 + |B|^2 \right) - \frac{1}{2} \left( AB^* + A^*B \right) = \frac{1}{2}, 
$$
which yields 
$$
A = B = \frac{e^{i\phi}}{\sqrt{2}},
$$
which is equivalent to the result in Eq.~({\ref{b8}) up to insignificant common phase $\phi$. 

It is easy to see that if we start from another linearly polarized product state $|0_x\rangle |1_y\rangle$ instead of Eq.~(\ref{b6}), we arrive at another Bell state: 
\begin{equation}
\label{b18}
\Psi_- = \frac{|1_R\rangle |0_L\rangle - |0_R\rangle |1_L\rangle }{\sqrt{2}}. 
\end{equation}

\end{document}